\documentclass[conference,letter]{IEEEtran}

\usepackage{preamble}

\usepackage{cite}
\usepackage{amsmath,amssymb,amsfonts}
\usepackage{algorithmic}
\usepackage{textcomp}
\usepackage{xcolor}
\usepackage{pifont}

\def\BibTeX{{\rm B\kern-.05em{\sc i\kern-.025em b}\kern-.08em
    T\kern-.1667em\lower.7ex\hbox{E}\kern-.125emX}}

\begin{document}

\title{Understanding the Ratio of the Partition Sum \\
       to its Bethe Approximation via Double Covers}

\author{
  \IEEEauthorblockN{Pascal O. Vontobel}
  \IEEEauthorblockA{
    \textit{Department of Information Engineering} \\
    \textit{The Chinese University of Hong Kong} \\
    pascal.vontobel@ieee.org \\[-0.5cm]
  }
}

\maketitle

\begin{abstract}
  For various classes of graphical models it has been observed that the ratio
  of the partition sum to its Bethe approximation is often close to being the
  square of the ratio of the partition sum to its degree-2 Bethe
  approximation. This is of relevance because the latter ratio can often
  better be analyzed and/or quantified than the former ratio. In this paper,
  we give some justifications for the observed relationship between these two
  ratios and then analyze these ratios for two classes of log-supermodular
  graphical models.
\end{abstract}

\section{Introduction}
\label{sec:introduction:1}

The partition sum (a.k.a.\ the partition function) of a normal factor graph
(NFG) $\graphN$~(see, e.g., \cite{Kschischang:Frey:Loeliger:01, Forney:01:1,
  Loeliger:04:1}) is defined to be
\begin{align}
  Z(\graphN)
    &\defeq
       \sum_{\va \in \setA}
         g(\va).
           \label{eq:partition:sum:1}
\end{align}
Here, the sum is over the set of all configurations $\setA$ of $\graphN$, and
$g$ is the global function of $\graphN$. The relevance of the partition sum
not only stems from the fact that it is an important quantity characterizing a
graphical model, but also from the fact that many quantities of interest can
be formulated as the partition sum of some suitable graphical model.

For many cases of interest, exactly computing the partition sum is
intractable. Therefore, various techniques have been proposed to efficiently
approximate the partition sum. For NFGs where the global function is a product
of non-negative-valued local functions, a popular approach is to approximate
$Z(\graphN)$ by the Bethe partition sum $\ZBethe(\graphN)$. The latter
quantity is defined via the minimum of the Bethe free energy
function~\cite{Yedidia:Freeman:Weiss:05:1} and for many graphical models it
can be found efficiently with the help of the sum-product algorithm
(SPA)~\cite{Kschischang:Frey:Loeliger:01, Loeliger:04:1,
  Yedidia:Freeman:Weiss:05:1}.

The above definition of $\ZBethe(\graphN)$ in terms of the minimum of the
Bethe free energy function can be considered to be an analytical definition of
$\ZBethe(\graphN)$. On the other hand, as was shown in~\cite{Vontobel:13:2},
$\ZBethe(\graphN)$ can also be given a combinatorial characterization as
follows:
\begin{align}
  \ZBethe(\graphN)
    &= \limsup_{M \to \infty} \ 
         \ZBetheM{M}(\graphN),
           \label{eq:ZBethe:intro:1} \\
  \ZBetheM{M}(\graphN)
    &\defeq
       \sqrt[M]{\Big\langle \!
                  Z(\cgraph{N})
                \! \Big\rangle_{\cgraph{N} \in \cset{N}_{M}}}.
                             \label{eq:ZBethe:intro:2}
\end{align}
Here the expression under the root sign represents the (arithmetic) average of
$Z(\cgraph{N})$ over all $M$-covers $\cgraph{N}$ of $\graphN$, $M \geq
1$. Note that for $M = 1$, the set $\cset{N}_{M}$ contains only the NFG
$\graphN$, and so
\begin{align}
  \ZBetheM{1}(\graphN)
    &= Z(\graphN).
         \label{eq:ZBethe:intro:3}
\end{align}
The results in~\eqref{eq:ZBethe:intro:1}--\eqref{eq:ZBethe:intro:3} are
summarized in Fig.~\ref{fig:degree:M:Bethe:partition:function:1}. More details
about the definition of NFGs and their graph covers will be given in
Section~\ref{sec:nfgs:and:their:covers:1}.

\begin{figure}
  {\footnotesize
  \begin{alignat*}{2}
    &\big.
       \ZBetheM{M}(\graphN)
     \big|_{M \to \infty}
        &&= \ZBethe(\graphN) \\
    &\hskip1cm \Big\vert \\
    &\big.
       \ZBetheM{M}(\graphN)
     \big. && \\
    &\hskip1cm \Big\vert \\
    &\big.
       \ZBetheM{M}(\graphN)
     \big|_{M = 1}
       &&= Z(\graphN)
  \end{alignat*}
  }
  \caption{The degree-$M$ Bethe partition sum of the NFG $\graphN$ for
    different values of $M$.}
  \label{fig:degree:M:Bethe:partition:function:1}
\end{figure}

For various classes of NFGs it has been numerically and/or analytically
observed that the following approximation holds:
\begin{align}
  \frac{Z(\graphN)}
       {\ZBethe(\graphN)}
    &\approx
       \left(
         \frac{Z(\graphN)}
              {\ZBetheM{2}(\graphN)}
       \right)^{\!\! 2}.
         \label{eq:approximate:Z:ZBethe:relationship:1}
\end{align}
In this paper, we would like to argue that this is likely a general
phenomenon. Note that~\eqref{eq:approximate:Z:ZBethe:relationship:1} can also
be rewritten as
\begin{align}
  \ZBetheM{2}(\graphN)
    &\approx 
       \sqrt{Z(\graphN) \cdot \ZBethe(\graphN)},
         \label{eq:approximate:Z:ZBethe:relationship:2}
\end{align}
i.e., $\ZBetheM{2}(\graphN)$ is approximately the geometric mean of the
quantities $Z(\graphN)$ and
$\ZBethe(\graphN)$. Eq.~\eqref{eq:approximate:Z:ZBethe:relationship:1} can
also be rewritten as
\begin{align}
  \rho(\graphN)
    &\defeq
       \frac{Z(\graphN) \cdot \ZBethe(\graphN)}
            {\bigl( \ZBetheM{2}(\graphN) \bigr)^2}
     \approx
       1.
         \label{eq:approximate:Z:ZBethe:relationship:3}
\end{align}
The expression in~\eqref{eq:approximate:Z:ZBethe:relationship:1} is of
relevance because, while the ratio on the LHS
of~\eqref{eq:approximate:Z:ZBethe:relationship:1} is of interest, the ratio on
the RHS of~\eqref{eq:approximate:Z:ZBethe:relationship:1} can often be better
analyzed and/or quantified.

Let us discuss a case where the approximate equality
in~\eqref{eq:approximate:Z:ZBethe:relationship:1} was observed. Namely, let
$\matr{A}$ be the all-one matrix of size $n \times n$ for an arbitrary
positive integer $n$. It was shown in~\cite{Vontobel:13:1} and
in~\cite{Ng:Vontobel:22:2}, respectively, that\footnote{The notation
  $a(n) \sim b(n)$ stands for
  $\lim\limits_{n \to \infty} \frac{a(n)}{b(n)} = 1$.}
\begin{align}
  \frac{\perm(\matr{A})}
       {\perm_{\Bethe}(\matr{A})}
    &\sim
       \sqrt{\frac{2\pi n}{\e}}, \\
  \frac{\perm(\matr{A})}
       {\perm_{\Bethe,2}(\matr{A})}
    &\sim
       \sqrt[4]{\frac{\pi n}{\e}}.
\end{align}
Observe that, up to a factor $\sqrt{2}$, the relationship
in~\eqref{eq:approximate:Z:ZBethe:relationship:1} is essentially satisfied for
large $n$.

\subsection{Overview}
\label{sec:overview:1}

The paper is structured as follows. In
Section~\ref{sec:nfgs:and:their:covers:1} we give a brief introduction to NFGs
and their double covers. In Section~\ref{sec:justification:1} we justify the
approximate equality in~\eqref{eq:approximate:Z:ZBethe:relationship:1}. In
Sections~\ref{sec:single:cycle:nfg:1}
and~\ref{sec:class:log:supermodular:nfg:1} we analyze $\rho(\graphN)$ for two
classes of NFGs, first a very simple class and then a more sophisticated class
consisting of log-supermodular graphical models. Finally, in
Section~\ref{sec:conclusion:1} we conclude the paper.

\section{Normal Factor Graphs and Their Finite Covers}
\label{sec:nfgs:and:their:covers:1}

Factor graphs are a convenient way to represent the factorization of
multivariate functions~\cite{Kschischang:Frey:Loeliger:01}. Normal factor
graphs (NFGs)~\cite{Forney:01:1}, also called Forney-style factor
graphs~\cite{Loeliger:04:1}, are a variant of factor graphs where variables
are associated with edges. The following example is taken
from~\cite{Vontobel:13:2}.

\begin{figure}
  \begin{center}
    \includegraphics[scale=0.88]
                    {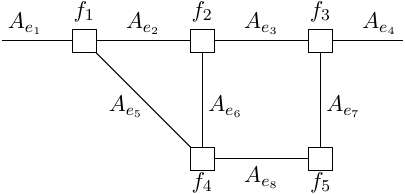}
  \end{center}
  \caption{NFG $\graphN$ used in Example~\ref{example:simple:ffg:1}.}
  \label{fig:example:simple:ffg:1}
\end{figure}

\begin{Example}
  \label{example:simple:ffg:1}

  Consider the multivariate function
  \begin{align*}
    g(a_{e_1}, \ldots, a_{e_8})
      &\defeq
         f_1(a_{e_1}, a_{e_2}, a_{e_5})
         \cdot
         f_2(a_{e_2}, a_{e_3}, a_{e_6}) \, \cdot \\
      &\hspace{-1cm}
         f_3(a_{e_3}, a_{e_4}, a_{e_7})
         \cdot
         f_4(a_{e_5}, a_{e_6}, a_{e_8})
         \cdot
         f_5(a_{e_7}, a_{e_8}),
  \end{align*}
  where the so-called global function $g$ is the product of the so-called
  local functions $f_1$, $f_2$, $f_3$, $f_4$, and $f_5$. The decomposition of
  this global function as a product of local functions can be depicted with
  the help of an NFG $\graphN$ as shown in
  Fig.~\ref{fig:example:simple:ffg:1}. In particular, the NFG $\graphN$
  consists of
  \begin{itemize}
    
  \item the function nodes $f_1$, $f_2$, $f_3$, $f_4$, and $f_5$;

  \item the half edges $e_1$ and $e_4$ (sometimes also called ``external
    edges'');

  \item the full edges $e_2$, $e_3$, $e_5$, $e_6$, $e_7$, and $e_8$ (sometimes
    also called ``internal edges'').

  \end{itemize}
  In general,
  \begin{itemize}

  \item a function node $f$ represents the local function $f$;

  \item with an edge $e$ we associate the variable $A_e$ (note that a
    realization of the variable $A_e$ is denoted by $a_e$);

  \item an edge $e$ is incident on a function node $f$ if and only if $a_e$
    appears as an argument of the local function $f$.

  \end{itemize}
  Finally, we associate with $\graphN$ the partition sum~$Z(\graphN)$ as
  defined in~\eqref{eq:partition:sum:1}. (Note that we do not consider any
  temperature dependence of $Z(\graphN)$ in this paper.)
\end{Example}

\begin{figure}
  \begin{center}
    \includegraphics[scale=0.88]
                    {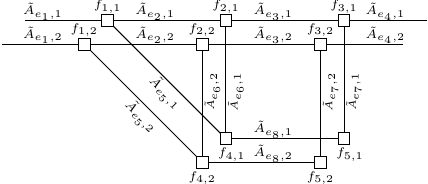} \\[0.5cm]
    \includegraphics[scale=0.88]
                    {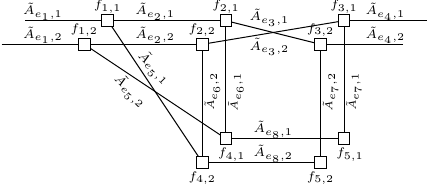}
  \end{center}
  \caption{Two possible $2$-covers of the NFG $\graphN$ in
    Fig.~\ref{fig:example:simple:ffg:1}.}
  \label{fig:example:simple:ffg:1:cover:1}
\end{figure}

Throughout this paper, we will essentially use the same notation as
in~\cite{Vontobel:13:2}. The only exceptions are $Z(\graphN)$ instead of
$\ZGibbs(\graphN)$ for the partition sum, and $f$ instead of $g_f$ for local
functions. (For notations which are not defined in this paper, we refer the
reader to~\cite{Vontobel:13:2}.) Note that for the rest of this paper, we
assume that local functions in the base NFG $\graphN$ take on only
non-negative real values, \ie, $f(\va_f) \in \R_{\geq 0}$ for all $f$ and all
$\va_f$. However, transformations of $\graphN$ might have local functions that
take on negative real values.

Central to this paper are also finite graph covers of an NFG. (For the
definition of a finite graph cover, see, \eg,~\cite{Vontobel:13:2}.) The
following example is taken from~\cite{Vontobel:13:2}.

\begin{Example}
  \label{example:simple:ffg:1:cont:1}

  Consider again the NFG $\graphN$ that is discussed in
  Example~\ref{example:simple:ffg:1} and depicted in
  Fig.~\ref{fig:example:simple:ffg:1}. Two possible $2$-covers of this
  (base) NFG are shown in Fig.~\ref{fig:example:simple:ffg:1:cover:1}. The
  first graph cover is ``trivial'' in the sense that it consists of two
  disjoint copies of the NFG in Fig.~\ref{fig:example:simple:ffg:1}. The
  second graph cover is ``more interesting'' in the sense that the edge
  permutations are such that the two copies of the base NFG are
  intertwined. (Of course, both graph covers are equally valid.)
\end{Example}

Based on finite graph covers, one can define the degree-$M$ Bethe partition
sum $\ZBetheM{M}(\graphN)$ as in~\eqref{eq:ZBethe:intro:2} for any $M \geq
1$. With this, one can prove the alternative expression for $\ZBethe(\graphN)$
in~\eqref{eq:ZBethe:intro:1}. When considering the value of
$\ZBetheM{M}(\graphN)$ from $M = 1$ to $M = \infty$, one goes from
$Z(\graphN)$ to $\ZBethe(\graphN)$ as shown in
Fig.~\ref{fig:degree:M:Bethe:partition:function:1}.

\section{Justification of the \\ Approximate Equality 
                in~\eqref{eq:approximate:Z:ZBethe:relationship:1}}
\label{sec:justification:1}

In this section, we would like to justify the approximate equality
in~\eqref{eq:approximate:Z:ZBethe:relationship:1}. Assume that the sequence
$\bigl\{ \ZBetheM{2^k}(\graphN) \bigr\}_{k=0}^{\infty}$ converges\footnote{We
  suspect that $\limsup_{M \to \infty}$ in~\eqref{eq:ZBethe:intro:1} can be
  replaced by $\lim_{M \to \infty}$ for general NFGs, but we are not aware of
  a proof of this.}  to $\ZBethe(\graphN)$. Then
\begin{align}
  \frac{Z(\graphN)}
       {\ZBethe(\graphN)}
    &= \frac{Z(\graphN)}
            {\ZBetheM{2}(\graphN)}
       \cdot
       \frac{\ZBetheM{2}(\graphN)}
            {\ZBetheM{4}(\graphN)}
       \cdot
       \frac{\ZBetheM{4}(\graphN)}
            {\ZBetheM{8}(\graphN)}
       \cdot
       \frac{\ZBetheM{8}(\graphN)}
            {\ZBetheM{16}(\graphN)}
       \cdots \ ,
         \label{eq:Z:ratios:2}
\end{align}
which holds thanks to the cancellation of terms on the RHS.

Let
\begin{align*}
  \eta 
    &\defeq
       \frac{Z(\graphN)}
            {\ZBetheM{2}(\graphN)}
     = \frac{\ZBetheM{1}(\graphN)}
            {\ZBetheM{2}(\graphN)},
\end{align*}
and assume that we can show that
\begin{align}
  \frac{\ZBetheM{2}(\graphN)}
       {\ZBetheM{4}(\graphN)}
    &\approx
       \eta^{1/2}, \quad
  \frac{\ZBetheM{4}(\graphN)}
       {\ZBetheM{8}(\graphN)}
     \approx
       \eta^{1/4}, \quad
  \ldots
    \label{eq:approximate:ZBethe:relationships:1}
\end{align}
Then Eq.~\eqref{eq:Z:ratios:2} can be rewritten as
\begin{align*}
  \frac{Z(\graphN)}
       {\ZBethe(\graphN)}
    &\approx
       \eta^{1/1}
       \cdot
       \eta^{1/2}
       \cdot
       \eta^{1/4}
       \cdot
       \eta^{1/8}
       \cdot
       \ \cdots
     = \eta^2
     = \left(
         \frac{Z(\graphN)}
              {\ZBetheM{2}(\graphN)}
       \right)^{\!\! 2},
\end{align*}
which is exactly the claimed relationship
in~\eqref{eq:approximate:Z:ZBethe:relationship:1}. Note that for for
$M = 2^k$, where $k \in \Z_{\geq 0}$,
Eq.~\eqref{eq:approximate:ZBethe:relationships:1} also implies
\begin{align}
  \frac{Z(\graphN)}
       {\ZBetheM{M}(\graphN)}
    &\approx
       \eta^{2 (1 - 1/M)}
     =
       \left(
         \frac{Z(\graphN)}
              {\ZBetheM{2}(\graphN)}
       \right)^{\!\! 2 (1 - 1/M)}.
\end{align}
Interestingly, the lower and upper bounds on
$Z(\graphN) / \ZBetheM{M}(\graphN)$ in Eq.~(10) of Theorem~4
in~\cite{Huang:Kashyap:Vontobel:24:1} are equal $\eta^{2 (1 - 1/M)}$ for
$\eta = 1$ and $\eta = 2^{n/4}$, respectively.

It remains to justify the expressions
in~\eqref{eq:approximate:ZBethe:relationships:1}. Indeed, the first expression
in~\eqref{eq:approximate:ZBethe:relationships:1} follows from taking the
fourth root on both sides of the equation
\begin{align*}
  \frac{\bigl( \ZBetheM{2}(\graphN) \bigr)^{\! 4}}
       {\bigl( \ZBetheM{4}(\graphN) \bigr)^{\! 4}}
    &= \frac
       {
         \Big\langle \!
           Z(\cgraph{N})
         \! \Big\rangle_{\cgraph{N} \in \cset{N}_{2}}^{\!\! 2}
       }
       {
         \Big\langle \!
           Z(\cgraph{N})
         \! \Big\rangle_{\cgraph{N} \in \cset{N}_{4}}
       }
     \approx
       \frac
       {
         \Big\langle \!
           Z(\cgraph{N})
         \! \Big\rangle_{\cgraph{N} \in \cset{N}_{1}}^{\!\! 2}
       }
       {
         \Big\langle \!
           Z(\cgraph{N})
         \! \Big\rangle_{\cgraph{N} \in \cset{N}_{2}}
       }
     = \frac{\bigl( \ZBetheM{1}(\graphN) \bigr)^{\! 2}}
            {\bigl( \ZBetheM{2}(\graphN) \bigr)^{\! 2}}.
\end{align*}
Here, the approximation sign is justified as follows: the way $4$-covers of
$\graphN$ behave to $2$-covers of $\graphN$ is similar to the way that
$2$-covers of $\graphN$ behave to $1$-covers of $\graphN$. (This is in general
not an equality as not all $4$-covers are $2$-covers of $2$-covers.)
Similarly, the second expression
in~\eqref{eq:approximate:ZBethe:relationships:1} follows from taking the
eighth root on both sides of the equation
\begin{align*}
  \frac{\bigl( \ZBetheM{4}(\graphN) \bigr)^{\! 8}}
       {\bigl( \ZBetheM{8}(\graphN) \bigr)^{\! 8}}
    &= \frac
       {
         \Big\langle \!
           Z(\cgraph{N})
         \! \Big\rangle_{\cgraph{N} \in \cset{N}_{4}}^{\!\! 2}
       }
       {
         \Big\langle \!
           Z(\cgraph{N})
         \! \Big\rangle_{\cgraph{N} \in \cset{N}_{8}}
       }
     \approx
       \frac
       {
         \Big\langle \!
           Z(\cgraph{N})
         \! \Big\rangle_{\cgraph{N} \in \cset{N}_{2}}^{\!\! 2}
       }
       {
         \Big\langle \!
           Z(\cgraph{N})
         \! \Big\rangle_{\cgraph{N} \in \cset{N}_{4}}
       }
     = \frac{\bigl( \ZBetheM{2}(\graphN) \bigr)^{\! 4}}
            {\bigl( \ZBetheM{4}(\graphN) \bigr)^{\! 4}}.
\end{align*}
Here, the approximation sign is justified as follows: the way $8$-covers of
$\graphN$ behave to $4$-covers of $\graphN$ is similar to the way that
$4$-covers of $\graphN$ behave to $2$-covers of $\graphN$. (This is in general
not an equality as not all $8$-covers are $2$-covers of $4$-covers.) The
justification for the remaining expressions
in~\eqref{eq:approximate:ZBethe:relationships:1} is similar.

This concludes the proposed justification of the approximate equality
in~\eqref{eq:approximate:Z:ZBethe:relationship:1}.

\section{Analysis of a Single-Cycle Graphical Model}
\label{sec:single:cycle:nfg:1}

\begin{figure}
  \begin{center}
    \includegraphics[scale=0.88]
                    {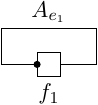}
  \end{center}
  \caption{NFG $\graphN$ discussed in Section~\ref{sec:single:cycle:nfg:1}.}
  \label{fig:single:cycle:nfg:1}
\end{figure}

Consider the NFG $\graphN$ in Fig.~\ref{fig:single:cycle:nfg:1} that consists
of a single function node $f_1$ of degree two and an edge $e_1$ with
associated variable $A_{e_1} \in \{ 0, 1 \}$. (The dot denotes the first
argument of $f_1$.) Let
\begin{align*}
  \matr{T}_{f_1}
    &\defeq 
       \begin{pmatrix}
         f_1(0,0) & f_1(0,1) \\
         f_1(1,0) & f_1(1,1)
       \end{pmatrix}.
\end{align*}
In order to keep the discussion simple, we assume that $f_1$ takes on strictly
positive values.

Let $\lambda_1$ and $\lambda_2$ with $|\lambda_1| \geq |\lambda_2|$ be the two
eigenvalues of $\matr{T}_{f_1}$. For the given setup we can use Perron-Frobenius
theory to conclude that $\lambda_1$ and $\lambda_2$ are real numbers
satisfying $\lambda_1 > |\lambda_2|$.

\begin{Proposition}
  \label{prop:partition:function:calculation:example:1}

  For the NFG $\graphN$ defined at the beginning of this section, it holds
  that
  \begin{align*}
    Z(\graphN)
      &= \sum_{a_{e_1}}
           f_1(a_{e_1}, a_{e_1})
       = \Traceshort(\matr{T}_{f_1})
       = \lambda_1 + \lambda_2, \\
    \ZBetheM{2}(\graphN)
      &= \sqrt
         {
           \frac{1}{2} \bigl( \Traceshort(\matr{T}_{f_1}) \bigr)^2
           \!\! + \!\!
           \frac{1}{2} \Traceshort\bigl( (\matr{T}_{f_1})^2 \bigr)
         }
       = \sqrt{\lambda_1^2 \!\! + \!\! \lambda_1 \lambda_2 \!\! + \!\! \lambda_2^2}, \\
    \ZBethe(\graphN)
      &= \lambda_1.
  \end{align*}
\end{Proposition}

\begin{Proof}
  See Appendix~\ref{app:proof:prop:partition:function:calculation:example:1}.
\end{Proof}

Computing the ratio on the LHS
of~\eqref{eq:approximate:Z:ZBethe:relationship:3}, we observe that
\begin{align*}
  \rho(\graphN)
    &= \frac{Z(\graphN) \cdot \ZBethe(\graphN)}
       {\bigl( \ZBetheM{2}(\graphN) \bigr)^2}
     = \frac{\lambda_1^2 + \lambda_1 \lambda_2}
            {\lambda_1^2 + \lambda_1 \lambda_2 + \lambda_2^2}
     = 1
       -
       \frac{\xi^2}
            {1 + \xi + \xi^2},
\end{align*}
where $\xi \defeq \lambda_2 / \lambda_1$. We see that the smaller $\xi$ is,
the closer the ratio on the LHS
of~\eqref{eq:approximate:Z:ZBethe:relationship:3} is to one. In particular, if
$|\xi + \xi^2| < 1$, then $\rho(\graphN) = 1 - \xi^2 + \xi^3 + O(\xi^4)$.

Let $\theta$ be a real number satisfying $0 < \theta \leq 1$. Consider the
special case where the function $f_1$ is such that
$\matr{T}_{f_1} \defeq \bigl( \begin{smallmatrix} 1 & \theta \\ \theta & 1
                              \end{smallmatrix} \bigr)$.
                              Note that the function $f_1$ is log-supermodular. (For details of the definition
                              of log-supermodularity, see, e.g., \cite{Ruozzi:12:1, Vontobel:16:1}.)

The eigenvalues of $\matr{T}_{f_1}$ are $\lambda_1 = 1 + \theta$ and
$\lambda_2 = 1 - \theta$, and so
$\xi = \frac{\lambda_2} {\lambda_1} = \frac{1 - \theta} {1 + \theta}$. We
observe that if $\theta$ is close to one, then $\xi$ is close to zero, and
with that the ratio $\rho(\graphN)$ is close to one. On the other hand, if
$\theta$ is close to zero, then $\xi$ is close to one, and with that the ratio
$\rho(\graphN)$ is close to $2/3$.

\section{Analysis of a Special Class of \\
                Log-supermodular Graphical Models}
\label{sec:class:log:supermodular:nfg:1}

\subsection{Definition of a Special Class of
                         Log-supermodular Graphical Models}
\label{sec:class:log:supermodular:nfg:1:definition:1}

Let $\theta$ be a real number satisfying $0 < \theta \leq 1$. Consider an NFG
$\graphN$ with $m$ function nodes and $n$ (full) edges, where all function
nodes have degree $3$ and represent the local function
\begin{align*}
  f_0(a_1, a_2, a_3)
    &\defeq
       \begin{cases}
         1      & \text{(if $a_1 = a_2 = a_3$)} \\
         \theta & \text{(otherwise)}
       \end{cases}
\end{align*}
Note:
\begin{itemize}

\item The parameters $m$ and $n$ have to satisfy $3m = 2n$.

\item The function $f_0$ is symmetric in the arguments. With this, the
  ordering of the arguments is irrelevant.

\item The function $f_0$ is log-supermodular. (For details of the definition of
  log-supermodularity, see, e.g., \cite{Ruozzi:12:1, Vontobel:16:1}.)

\end{itemize}

\subsection{The Bethe Approximation of the Partition Sum}
\label{sec:class:log:supermodular:nfg:1:Bethe:1}

Let us determine $\ZBethe(\graphN)$ by finding the sum-product algorithm (SPA)
fixed-point messages~\cite{Yedidia:Freeman:Weiss:05:1}. Because of the
symmetry of the setup, we may assume that all the messages along all the edges
in all directions are the same, say
$\mu = \bigl( \mu(0), \mu(1) \bigr) = \frac{1}{\Lambda + 1} \cdot ( \Lambda, 1
)$ for some $\Lambda \in \R_{\geq 0} \cup \{ + \infty \}$.

\begin{Lemma}
  \label{lemma:SPA:fixed:point:messages:1}
 
  If $1/5 \leq \theta \leq 1$, then the SPA has one fixed-point message
  corresponding to
  \begin{align*}
    \Lambda_0
      &= 1.
  \end{align*}
  Note that $\Lambda_0$ corresponds to a stable SPA fixed-point.

  If $0 < \theta < 1/5$, then the SPA has three fixed-point messages
  corresponding to
  \begin{align*}
    \Lambda_0
      &= 1, \quad
    \Lambda_{\pm}
       = \frac{1}
              {2\theta}
           \cdot
           \bigl( 1 - 3\theta \pm \sqrt{5\theta^2 - 6\theta + 1} \bigr).
  \end{align*}
  Note that $\Lambda_{-} = \Lambda_{+}^{-1}$. Moreover, $\Lambda_{+}$ and
  $\Lambda_{-}$ correspond to stable SPA fixed points, whereas $\Lambda_0$
  corresponds to an unstable SPA fixed point.

\end{Lemma}

\begin{Proof}
  See Appendix~\ref{app:proof:lemma:SPA:fixed:point:messages:1}.
\end{Proof}

\begin{Remark}
  It follows from Lemma~\ref{lemma:SPA:fixed:point:messages:1} that the Bethe
  approximation for the class of NFGs under consideration has a phase
  transition at $\theta^* = 1/5$.
\end{Remark}

\begin{Proposition}
  If $1/5 \leq \theta \leq 1$, then it holds that
  \begin{align*}
    \ZBethe(\graphN)
      &= \frac{(2 + 6\theta)^m}
              {2^n}
       = \left(
           \frac{2 + 6\theta}
                {\sqrt{8}}
         \right)^{\!\! m}.
  \end{align*}
  (We omit the corresponding expression for $\ZBethe$ for the case
  $0 < \theta < 1/5$.)
\end{Proposition}

\begin{Proof}
  The Bethe partition sum is given by
  \begin{align*}
    \ZBethe(\graphN)
      &= \frac{\prod_f Z_f}
              {\prod_e Z_e},
  \end{align*}
  where the product in the numerator is over all function nodes, where the
  product in the denominator is over all full edges, and where, thanks to the
  fact that all the SPA fixed-point messages are the same, the expressions for
  $Z_f$ and $Z_e$ are given by
  \begin{align*}
    Z_f
      &= \sum_{a_1,a_2,a_3}
           f_0(a_1,a_2,a_3) \mu(a_1) \mu(a_2) \mu(a_3), \\
    Z_e
      &= \bigl( \mu(0) \bigr)^2
         +
         \bigl( \mu(1) \bigr)^2.
  \end{align*}
  Using the results from Lemma~\ref{lemma:SPA:fixed:point:messages:1}, we
  obtain the result in the proposition statement.
\end{Proof}

\subsection{Loop-Calculus Transform}
\label{sec:class:log:supermodular:nfg:1:LCT:1}

It turns out that the loop-calculus approach by Chertkov and
Chernyak~\cite{Chertkov:Chernyak:06:1} can be beneficially used to better
understand the connection of $\ZBethe(\graphN)$ and $Z(\graphN)$ for the class
of NFGs under consideration in this section. Here we apply the loop-calculus
approach in terms of the loop-calculus transform (LCT) as defined
in\cite[Section~VI.B]{Forney:Vontobel:11:1} and obtain a new NFG $\graphNLCT$
with global function $\gLCT$. Let $\fzLCT$ be the function node in
$\graphNLCT$ corresponding to $f_0$.

\begin{Lemma}
  \label{lemma:f:LCT:1}

  If $1/5 \leq \theta \leq 1$, then the LCT based on $\Lambda = \Lambda_0$
  yields \\[-0.75cm]
  \begin{align*}
    \fzLCT(a_1,a_2,a_3)
      &= \begin{cases}
           \frac{2 + 6\theta}{\sqrt{8}} 
             & \text{(if $(a_1,a_2,a_3) = (0,0,0)$)} \\
           \frac{2 - 2\theta}{\sqrt{8}} 
             & \text{(if $(a_1,a_2,a_3) = (0,1,1)$)} \\
           \frac{2 - 2\theta}{\sqrt{8}} 
             & \text{(if $(a_1,a_2,a_3) = (1,0,1)$)} \\
           \frac{2 - 2\theta}{\sqrt{8}} 
             & \text{(if $(a_1,a_2,a_3) = (1,1,0)$)} \\
           0
             & \text{(otherwise)} \\
         \end{cases}
  \end{align*}
  (We omit the corresponding expression for $0 < \theta < 1/5$ that is
  obtained by an LCT based on $\Lambda = \Lambda_{+}$, but we note that
  $\fzLCT(a_1,a_2,a_3) > 0$ for $(a_1,a_2,a_3) = (1,1,1)$.)
\end{Lemma}

\begin{Proof}
  Omitted.
\end{Proof}

Note that, by design of the LCT:
\begin{itemize}

\item
  $\ZBethe(\graphN) = \gLCT(\vect{0}) = \prod_f \fzLCT(0,0,0) \overset{(*)}{=}
  \left( \frac{2 + 6\theta} {\sqrt{8}} \right)^{\!\! m}$, where $(*)$ is for
  the case $1/5 \leq \theta \leq 1$.

\item
  $Z(\graphN) = \sum_{\va} \gLCT(\va) = \ZBethe(\graphN) + \sum_{\va \neq \vect{0}}
  \gLCT(\va)$.

\item $\fzLCT(a_1,a_2,a_3) = 0$ for
  $(a_1,a_2,a_3) = (1,0,0)$, \ $(0,1,0)$, \ $(0,0,1)$.

\end{itemize}

\begin{Remark}
  \label{remark:lcd:valid:configurations:1}

  Assume $1/5 \leq \theta \leq 1$. It follows from Lemma~\ref{lemma:f:LCT:1}
  that the valid configurations of $\graphNLCT$ have a simple structure: the
  edges that take on the value $1$ (more precisely, the edges whose associated
  variable takes on the value $1$), form a cycle or a disjoint union of
  cycles. In fact, there is a bijection between the set of valid
  configurations of $\graphNLCT$ and the set of codewords of a $(2,3)$-regular
  LDPC code $\setCLCT$ whose Tanner graph is represented by $\graphNLCT$ after
  inserting a variable node in every edge.\footnote{Because of their codeword
    structure, such codes are also known as cycle codes. (This is not to be
    confused with cyclic codes.)}
\end{Remark}

\begin{Lemma}
  \label{lemma:gLCT:value:1}

  Let $\va$ be a valid configuration of $\graphNLCT$, i.e.,
  $\va \in \setCLCT$. Then it holds that
  \begin{align*}
    \gLCT(\va)
      &= \ZBethe(\graphN)
         \cdot
         \left(
           \frac{2 - 2\theta}
                {2 + 6\theta}
         \right)^{\!\! \wH(\va)},
  \end{align*} 
  where $\wH(\va)$ is the Hamming weight of $\va$.
\end{Lemma}

\begin{Proof}
  This result follows from the observation that a cycle visits equally many
  edges as function nodes. Similarly, a disjoint union of cycles visits
  equally many edges as function nodes.
\end{Proof}

Note that for $1/5 \leq \theta \leq 1$ it holds that
$\frac{2 - 2\theta}{2 + 6\theta} \leq 1/2$. It follows from
Lemma~\ref{lemma:gLCT:value:1} that the larger $\wH(\va)$ is, the smaller is
the contribution of $\gLCT(\va)$ to $\sum_{\va} \gLCT(\va)$. Moreover,
\begin{align*}
  Z(\graphN) 
    &= \ZBethe(\graphN)
       +
       \sum_{\va \in \setCLCT \setminus \{ \vect{0} \}} \gLCT(\va) \\
    &= \ZBethe(\graphN)
       \cdot
       \Biggl(
         1
         +
         \underbrace{
           \sum_{\va \in \setCLCT \setminus \{ \vect{0} \}}
             \left(
               \frac{2 - 2\theta}
                    {2 + 6\theta}
             \right)^{\!\! \wH(\va)}
         }_{(*)}
       \Biggr).
\end{align*}
Using well-known techniques for studying the expected Hamming weight
distribution of regular LDPC codes (see, e.g., \cite{Richardson:Urbanke:08:1};
details are omitted), one can conclude that the expression in $(*)$ is
typically at most on the order of $1$, and so $\ZBethe(\graphN)$ gives a
(very) good approximation of $Z(\graphN)$.

For the case $0 < \theta < 1/5$, similar conclusions can be reached, but the
analysis is more elaborate as the set of valid configurations is larger.

\subsection{Double-Cover Transform}
\label{sec:dct:1}

In order to find $\ZBetheM{2}(\graphN)$, we apply the double-cover analysis
technique in~\cite{Vontobel:16:1}. Actually, we apply the technique
in~\cite{Vontobel:16:1} not to $\graphN$, but to $\graphNLCT$, resulting in a
graphical model called $\graphNLCTDCT$ with global function $\gLCTDCT$ and
variables that take value in
$\bigl\{ (0,0), \ (0,1), \ (1,0), \ (1,1) \bigr\}$. 
(Some details of the
relevant calculations are given
in Appendix~\ref{app:details:calculations:dct:1}.)

For simplicity, we consider only the case $1/5 \leq \theta \leq 1$ in the
following. We obtain
\begin{align*}
  \bigl( \ZBetheM{2}(\graphN) \bigr)^2
    &= \bigl( \ZBethe(\graphN) \bigr)^2
       \cdot
       \Biggl(
         1
         + \!\!\!\!\!\!
         \underbrace{
           \sum_{\hva \in \setCLCTDCT \setminus \{ \vect{0} \}} \!\!\!\!\!\!
             \frac{\gLCTDCT(\hva)}
                  {\bigl( \ZBethe(\graphN) \bigr)^2}
         }_{(**)}
       \Biggr). \\[-0.75cm]
\end{align*}
Compare this with
\begin{align*}
  Z(\graphN) \cdot \ZBethe(\graphN)
    &= \bigl( \ZBethe(\graphN) \bigr)^2
       \cdot
       \Biggl(
         1
         +
         \underbrace{
           \sum_{\va \in \setCLCT \setminus \{ \vect{0} \}}
             \frac{\gLCT(\va)}
                  {\ZBethe(\graphN)}
         }_{(***)}
       \Biggr).
\end{align*}
It turns out (details are omitted) that the dominating contribution to $(**)$
come from valid configurations that correspond to $(0,1)$-cycles and disjoint
unions of $(0,1)$-cycles. Moreover, there is not only a bijection between
these valid configurations in $(**)$ and the valid configurations in
$({*}{*}{*})$, but also their global function value (normalized by
$\bigl( \ZBethe(\graphN) \bigr)^2$ and $\ZBethe(\graphN)$, respectively) is
the same. With this, the relationship
\begin{align*}
  Z(\graphN) \cdot \ZBethe(\graphN)
    &\approx
       \bigl( \ZBetheM{2}(\graphN) \bigr)^2
\end{align*}
is not too surprising for the NFG under consideration.

\subsection{Numerical Results}
\label{sec:numerical:results:1}

\begin{figure}
  \begin{center}
    \includegraphics[width=0.95\linewidth]
                    {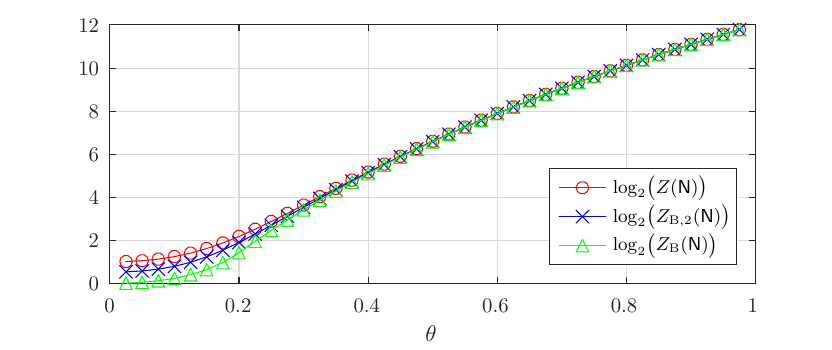}
  \end{center}
  \caption{Numerical results for NFG $\graphN$ discussed in
    Section~\ref{sec:numerical:results:1}.}
  \label{fig:example:log:supermodular:nfg:1:1}
\end{figure}

\begin{figure}
  \begin{center}
    \includegraphics[width=0.95\linewidth]
                    {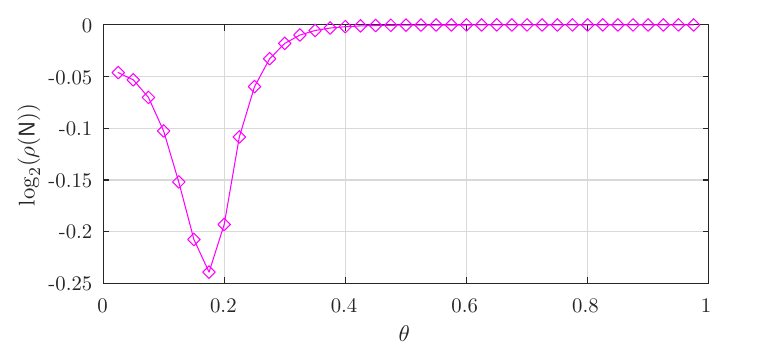}
  \end{center}
  \caption{Numerical results for NFG $\graphN$ discussed in
    Section~\ref{sec:numerical:results:1}.}
  \label{fig:example:log:supermodular:nfg:1:2}
\end{figure}

We consider a particular instance of the class of graphical models under
consideration in this section. Namely, let $\graphN$ have $m = 8$ function
nodes, $n = 12$ edges, and let its graph structure be described by the
(randomly generated) incidence matrix $\matr{H}$ shown
in Appendix~\ref{app:incidence:matrix:1},
where the $8$ rows and $12$ columns of
$\matr{H}$ correspond to, respectively, the $8$ function nodes and the $12$
edges of $\graphN$. Note that $\matr{H}$ happens to be the parity-check matrix
of the $(2,3)$-regular LDPC code
$\setCLCT$. Fig.~\ref{fig:example:log:supermodular:nfg:1:1} shows
$\log_2\bigl( Z(\graphN) \bigr)$, $\log_2\bigl( \ZBetheM{2}(\graphN) \bigr)$,
and $\log_2\bigl( \ZBethe(\graphN) \bigr)$ as a function of the parameter
$\theta$. We make the following observations:
\begin{itemize}

\item For $0.4 \lessapprox \theta \leq 1$, the Bethe approximation
  $\ZBethe(\graphN)$ is very close to the partition sum $Z(\graphN)$.

\item For all values of $\theta$, the blue curve for
  $\log\bigl( \ZBetheM{2}(\graphN) \bigr)$ is roughly half the way between the
  red curve for $\log\bigl( Z(\graphN) \bigr)$ and the green curve for
  $\log\bigl( \ZBethe(\graphN) \bigr)$, thereby showing that
  $\rho(\graphN) \approx 1$, where $\rho(\graphN)$ was defined
  in~\eqref{eq:approximate:Z:ZBethe:relationship:3}.

\end{itemize}
Fig.~\ref{fig:example:log:supermodular:nfg:1:2} considers the value of
$\log_2\bigl( \rho(\graphN) \bigr)$ as a function of $\theta$. We make the
following observations:
\begin{itemize}

\item For $0.4 \lessapprox \theta \leq 1$, the ratio $\rho(\graphN)$ is very
  close to one.

\item Interestingly, the ratio $\rho(\graphN)$ is the furthest from $1$ when
  $\theta$ is close to the phase transition point $\theta^* = 1/5$. (Note that
  $2^{-0.25} \approx 0.841$.)

\end{itemize}
Finally, note that Ruozzi~\cite{Ruozzi:12:1} showed that for any
log-supermodular NFG $\graphN$, any positive integer $M$, and any
$\cgraph{N} \in \cset{N}_{M}$, it holds that
$Z(\cgraph{N}) \leq Z(\graphN)^M$. For the class of log-supermodular NFGs
under consideration in this section we can significantly strengthen this
result for $M = 2$. Namely, for most $\cgraph{N} \in \cset{N}_{2}$ it holds
that $Z(\cgraph{N})$ is very close to $Z(\graphN)^2$.

\section{Conclusion}
\label{sec:conclusion:1}

A first key result of this paper is the approximate equality
in~\eqref{eq:approximate:Z:ZBethe:relationship:1} and its justification in
Section~\ref{sec:justification:1}. It remains to be seen to what extent the
approximation in this equation can be turned into a rigorous mathematical
statement, at least for some classes of NFGs.

Another key result of this paper was to show how the loop-calculus transform
(LCT) and the double-cover transform (DCT) can be combined to yield insightful
results about the relationship of $Z(\graphN)$, $\ZBetheM{2}(\graphN)$, and
$\ZBethe(\graphN)$ in general, and the ratio $\rho(\graphN)$ in particular,
for the specific class of log-supermodular NFGs under consideration in
Section~\ref{sec:class:log:supermodular:nfg:1}.

It remains to be seen what techniques can be used to obtain results about
$Z(\graphN)$, $\ZBetheM{2}(\graphN)$, $\ZBethe(\graphN)$, and $\rho(\graphN)$
for other classes of NFGs, e.g., the NFGs in~\cite{Straszak:Vishnoi:19:1,
  Csikvari:Ruozzi:Shams:22:1, Huang:Vontobel:24:1}.

\noindent 
{\scriptsize The research in this paper was partially supported by
  CUHK Direct Grant No.~4055204.}

\newpage

\appendices

\section{Proof of 
               Proposition~\ref{prop:partition:function:calculation:example:1}}
\label{app:proof:prop:partition:function:calculation:example:1}

The result for $Z(\graphN)$ is straightforward.

The result for $\ZBetheM{2}(\graphN)$ follows from the observation that the
NFG $\graphN$ in Fig.~\ref{fig:single:cycle:nfg:1} has two double covers:
\begin{itemize}

\item One double cover consists of two independent copies of $\graphN$,
  resulting in a contribution of $\bigl( \Traceshort(\matr{T}_{f_1}) \bigr)^2$
  to $\bigl( \ZBetheM{2}(\graphN) \bigr)^2$.

\item One double cover consists of a single cycle where the function node
  $f_1$ appears twice in sequence, resulting in a contribution of
  $\Traceshort\bigl( (\matr{T}_{f_1})^2 \bigr)$ to
  $\bigl( \ZBetheM{2}(\graphN) \bigr)^2$.

\end{itemize}

The result for $\ZBethe(\graphN)$ follows from the observation that the right
Perron eigenvector of $\matr{T}_{f_1}$ forms an SPA fixed-point message for
the message going in the clockwise direction of the cycle in the NFG $\graphN$
in Fig.~\ref{fig:single:cycle:nfg:1}, whereas the left Perron eigenvector of
$\matr{T}_{f_1}$ forms an SPA fixed-point message for the message going in the
counter-clockwise direction of the cycle in the NFG $\graphN$ in
Fig.~\ref{fig:single:cycle:nfg:1}.

\section{Proof of Lemma~\ref{lemma:SPA:fixed:point:messages:1}}
  \label{app:proof:lemma:SPA:fixed:point:messages:1}

  From the definition of $f_0$, it follows that the SPA fixed-point messages
  have to satisfy
  \begin{align*}
    \frac{\mu(0)}
         {\mu(1)}
      &= \frac{\begin{array}{c}
                   f_0(0,0,0) \mu(0) \mu(0) + f_0(0,0,1) \mu(0) \mu(1)\\
                 + f_0(0,1,0) \mu(1) \mu(0) + f_0(0,1,1) \mu(1) \mu(1)
               \end{array}}
              {\begin{array}{c}
                   f_0(1,0,0) \mu(0) \mu(0) + f_0(1,0,1) \mu(0) \mu(1) \\
                 + f_0(1,1,0) \mu(1) \mu(0) + f_0(1,1,1) \mu(1) \mu(1)
               \end{array}}.
  \end{align*}
  Using the definitions of $\Lambda$ and $f_0$, this equation becomes
  \begin{align*}
    \Lambda
      &= \frac{\Lambda^2 + 2 \theta \Lambda + \theta}
              {\theta \Lambda^2 + 2 \theta \Lambda + 1},
  \end{align*}
  which can be rewritten as
  \begin{align}
    \Lambda^3
    +
    (2 \! - \! \theta^{-1}) \Lambda^2
    +
    (\theta^{-1} \! - \! 2) \Lambda
    -
    1
      &= 0.
            \label{eq:SPA:fixed:point:Lambda:1}
  \end{align}
  It turns out that $\Lambda = 1$ is always a root
  of~\eqref{eq:SPA:fixed:point:Lambda:1}. With this,
  Eq.~\eqref{eq:SPA:fixed:point:Lambda:1} can be rewritten as
  \begin{align*}
    (\Lambda - 1)
    \cdot
    \underbrace{
      \bigl(
        \Lambda^2
        +
        (3 \! - \! \theta^{-1}) \Lambda
        +
        1
      \bigr)
    }_{(*)}
      &= 0.
  \end{align*}
  The above observations allow us to draw the following conclusions.
  \begin{itemize}

  \item If $0 < \theta < 1/5$, then the discriminant of the quadratic
    polynomial in $(*)$ is positive and we obtain the solutions mentioned in
    the lemma statement.

  \item If $\theta = 1/5$, then the discriminant of the quadratic polynomial
    in $(*)$ in zero. Given that $\Lambda = 1$ is a zero of multiplicity two
    of $(*)$, it turns out that $\Lambda = 1$ is a triple root
    of~\eqref{eq:SPA:fixed:point:Lambda:1}.

  \item If $1/5 < \theta \leq 1$, then the discriminant of the quadratic
    polynomial in $(*)$ is negative and $\Lambda = 1$ is the only real root
    of~\eqref{eq:SPA:fixed:point:Lambda:1}.

  \end{itemize}

\newpage

\section{Some Details of the Calculations in Section~\ref{sec:dct:1}}
\label{app:details:calculations:dct:1}

The technique that we are using is explained in Sections~III and~IV
of~\cite{Vontobel:16:1}; we refer to that paper for all the details. Here we
will only show the key results from Section IV.B of~\cite{Vontobel:16:1}.

Namely, for a local function $f$ with three arguments that take value in
$\{ 0, 1 \}$, the analysis technique in~\cite{Vontobel:16:1} introduces a
local function $\hat{f}$ with three arguments that take value in
$\bigl\{ (0,0), \ (0,1), \ (1,0), \ (1,1) \bigr\}$. The $4 \times 4 \times 4$
array $\matr{T}_{\hat{f}}$ associated with $\hat{f}$ is given by

{\scriptsize

\begin{center}
  $\displaystyle
  \left(
    \begin{array}{C{1.70cm}|C{1.70cm}C{1.70cm}|C{1.70cm}}
      t_{000} t_{000} 
        & \sqrt{2} \, t_{000} t_{010}
        & 0
        & t_{010} t_{010} \\[0.08cm]
    \hline
     & & & \\[-0.25cm]
      \sqrt{2} \, t_{000} t_{100} 
        & \perm(\matr{T}_{f|a_3=0})
        & 0 
        & \sqrt{2} \, t_{010} t_{110} \\
      0 
        & 0 
        & \det(\matr{T}_{f|a_3=0}) 
        & 0 \\[0.08cm]
    \hline
    & & & \\[-0.25cm]
      t_{100} t_{100} 
        & \sqrt{2} \, t_{100} t_{110} 
        & 0 
        & t_{110} t_{110}
    \end{array}
  \right)$, \\[0.35cm]
  $\displaystyle
  \left(
    \begin{array}{C{1.70cm}|C{1.70cm}C{1.70cm}|C{1.70cm}}
      \sqrt{2} \, t_{000} t_{001} 
        & \perm(\matr{T}_{f|a_1=0}) 
        & 0 
        & \sqrt{2} \, t_{010} t_{011} \\[0.08cm]
    \hline
    & & & \\[-0.25cm]
      \perm(\matr{T}_{f|a_2=0}) 
        & \hcf\big( \hzero, \hzero, \hzero \big)
        & 0 
        & \perm(\matr{T}_{f|a_2=1})  \\
      0 
        & 0 
        & \hcf\big( \hone, \hone, \hzero \big)
        & 0 \\[0.08cm]
    \hline
    & & & \\[-0.25cm]
      \sqrt{2} \, t_{100} t_{101} 
        & \perm(\matr{T}_{f|a_1=1}) 
        & 0 
        & \sqrt{2} \, t_{110} t_{111}
    \end{array}
  \right)$, \\[0.35cm]
  $\displaystyle
  \left(
    \begin{array}{C{1.70cm}|C{1.70cm}C{1.70cm}|C{1.70cm}}
      0 
        & 0 
        & \det(\matr{T}_{f|a_1=0}) 
        & 0 \\[0.08cm]
    \hline
    & & & \\[-0.25cm]
      0 
        & 0 
        & \hcf\big( \hzero, \hone, \hone \big)
        & 0 \\
      \det(\matr{T}_{f|a_2=0}) 
        & \hcf\big( \hone, \hzero, \hone \big)
        & 0 
        & \det(\matr{T}_{f|a_2=1}) \\[0.08cm]
    \hline
    & & & \\[-0.25cm]
      0 
        & 0 
        & \det(\matr{T}_{f|a_1=1}) 
        & 0
    \end{array}
  \right)$, \\[0.35cm]
  $\displaystyle
  \left(
    \begin{array}{C{1.70cm}|C{1.70cm}C{1.70cm}|C{1.70cm}}
      t_{001} t_{001} 
        & \sqrt{2} \, t_{001} t_{011} 
        & 0 
        & t_{011} t_{011} \\[0.08cm]
    \hline
    & & & \\[-0.25cm]
      \sqrt{2} \, t_{001} t_{101} 
        & \perm(\matr{T}_{f|a_3=1}) 
        & 0 
        & \sqrt{2} \, t_{011} t_{111} \\
      0 
        & 0 
        & \det(\matr{T}_{f|a_3=1}) 
        & 0 \\[0.08cm]
    \hline
    & & & \\[-0.25cm]
      t_{101} t_{101} 
        & \sqrt{2} \, t_{101} t_{111} 
        & 0 
        & t_{111} t_{111}
    \end{array}
  \right)$. \\[0.35cm]
\end{center}

}

\noindent
Here:
\begin{itemize}

\item The rows of the above matrices are indexed by
  $\bigl\{ (0,0), \ (0,1), \ (1,0), \ (1,1) \bigr\}$ and correspond to the
  values taken by the first argument of $\hat{f}$.

\item The columns of the above matrices are indexed by
  $\bigl\{ (0,0), \ (0,1), \ (1,0), \ (1,1) \bigr\}$ and correspond to the
  values taken by the second argument of $\hat{f}$.

\item The above four matrices are indexed by
  $\bigl\{ (0,0), \ (0,1), \ (1,0), \ (1,1) \bigr\}$ and correspond to the
  values taken by the third argument of $\hat{f}$.

\item $t_{000} \defeq f(0,0,0)$, $t_{001} \defeq f(0,0,1)$, \etc

\item $\matr{T}_{f|a_3=0}$ and $\matr{T}_{f|a_3=1}$ are the matrices
  associated with the functions $f(a_1,a_2,0)$ and $f(a_1,a_2,1)$,
  respectively. The matrices $\matr{T}_{f|a_1=0}$, $\matr{T}_{f|a_1=1}$,
  $\matr{T}_{f|a_2=0}$, and $\matr{T}_{f|a_2=1}$ are defined analogously.

\item $\hzero \defeq (0,1)$, $\hone \defeq (1,0)$, and $\gamma \defeq 1 /
\sqrt{2}$.

\item 
  $\hcf(\hzero,\hzero,\hzero)
     = \gamma
       \cdot
       (
         t_{000} t_{111} 
         + 
         t_{100} t_{011} 
         +
         t_{010} t_{101}
         +
         t_{001} t_{110}
       )$.

\item
  $\hcf(\hone,\hzero,\hone)
     = \gamma
       \cdot
       (
         t_{000} t_{111} 
         -
         t_{100} t_{011} 
         +
         t_{010} t_{101}
         -
         t_{001} t_{110}
       )$.

\item 
  $\hcf(\hzero,\hone,\hone)
     = \gamma
       \cdot
       (
         t_{000} t_{111} 
         +
         t_{100} t_{011} 
         -
         t_{010} t_{101}
         -
         t_{001} t_{110}
       )$.

\item
  $\hcf(\hone,\hone,\hzero)
    = \gamma
       \cdot
       (
         t_{000} t_{111} 
         - 
         t_{100} t_{011} 
         -
         t_{010} t_{101}
         +
         t_{001} t_{110}
       )$.

\end{itemize}
(The above $4 \times 4 \times 4$ array corrects a few minor typos in the
corresponding array in~\cite{Vontobel:16:1}.)

\bigskip

(continued on the next page)

\newpage

The above result was for an arbitrary local function $f$ with three arguments
each taking values in $\{ 0, 1 \}$. If $f$ is the result of an LCT, then
$f(1,0,0) = t_{100} = 0$, $f(0,1,0) = t_{010} = 0$, $f(0,0,1) = t_{001} =
0$, and the above $4 \times 4 \times 4$
array $\matr{T}_{\hat{f}}$ simplifies to

{\scriptsize

\begin{center}
  $\displaystyle
  \left(
    \begin{array}{C{1.70cm}|C{1.70cm}C{1.70cm}|C{1.70cm}}
      t_{000} t_{000} 
        & 0
        & 0
        & 0 \\[0.08cm]
    \hline
     & & & \\[-0.25cm]
      0 
        & t_{000} t_{110}
        & 0 
        & 0 \\
      0 
        & 0 
        & t_{000} t_{110}
        & 0 \\[0.08cm]
    \hline
    & & & \\[-0.25cm]
      0
        & 0
        & 0 
        & t_{110} t_{110}
    \end{array}
  \right)$, \\[0.35cm]
  $\displaystyle
  \left(
    \begin{array}{C{1.70cm}|C{1.70cm}C{1.70cm}|C{1.70cm}}
      0
        & t_{000} t_{011}
        & 0 
        & 0 \\[0.08cm]
    \hline
    & & & \\[-0.25cm]
      t_{000} t_{101}
        & \sqrt{2}^{-1} t_{000} t_{111}
        & 0 
        & t_{011} t_{110}  \\
      0 
        & 0 
        & \sqrt{2}^{-1} t_{000} t_{111}
        & 0 \\[0.08cm]
    \hline
    & & & \\[-0.25cm]
      0
        & t_{101} t_{110}
        & 0 
        & \sqrt{2} \, t_{110} t_{111}
    \end{array}
  \right)$, \\[0.35cm]
  $\displaystyle
  \left(
    \begin{array}{C{1.70cm}|C{1.70cm}C{1.70cm}|C{1.70cm}}
      0 
        & 0 
        & t_{000} t_{011} 
        & 0 \\[0.08cm]
    \hline
    & & & \\[-0.25cm]
      0 
        & 0 
        & \sqrt{2}^{-1} t_{000} t_{111}
        & 0 \\
      t_{000} t_{101}
        & \sqrt{2}^{-1} t_{000} t_{111}
        & 0 
        & - t_{011} t_{110} \\[0.08cm]
    \hline
    & & & \\[-0.25cm]
      0 
        & 0 
        & - t_{101} t_{110}
        & 0
    \end{array}
  \right)$, \\[0.35cm]
  $\displaystyle
  \left(
    \begin{array}{C{1.70cm}|C{1.70cm}C{1.70cm}|C{1.70cm}}
      0
        & 0
        & 0 
        & t_{011} t_{011} \\[0.08cm]
    \hline
    & & & \\[-0.25cm]
      0
        & t_{011} t_{101}
        & 0 
        & \sqrt{2} \, t_{011} t_{111} \\
      0 
        & 0 
        & - t_{011} t_{101}
        & 0 \\[0.08cm]
    \hline
    & & & \\[-0.25cm]
      t_{101} t_{101} 
        & \sqrt{2} \, t_{101} t_{111} 
        & 0 
        & t_{111} t_{111}
    \end{array}
  \right)$. \\[0.35cm]
\end{center}

}

It follows from the derivations in~\cite{Vontobel:16:1} that function values
of $\hat{f}$ where at least one of the arguments equals $(1,0)$ do not appear
in the calculation of $\ZBetheM{2}(\graphN)$. With this, the above
$4 \times 4 \times 4$ array $\matr{T}_{\hat{f}}$ can be further simplified to

{\scriptsize

\begin{center}
  $\displaystyle
  \left(
    \begin{array}{C{1.70cm}|C{1.70cm}C{1.70cm}|C{1.70cm}}
      t_{000} t_{000} 
        & 0
        & 0
        & 0 \\[0.08cm]
    \hline
     & & & \\[-0.25cm]
      0 
        & t_{000} t_{110}
        & 0 
        & 0 \\
      0 
        & 0 
        & 0
        & 0 \\[0.08cm]
    \hline
    & & & \\[-0.25cm]
      0
        & 0
        & 0 
        & t_{110} t_{110}
    \end{array}
  \right)$, \\[0.35cm]
  $\displaystyle
  \left(
    \begin{array}{C{1.70cm}|C{1.70cm}C{1.70cm}|C{1.70cm}}
      0
        & t_{000} t_{011}
        & 0 
        & 0 \\[0.08cm]
    \hline
    & & & \\[-0.25cm]
      t_{000} t_{101}
        & \sqrt{2}^{-1} t_{000} t_{111}
        & 0 
        & t_{011} t_{110}  \\
      0 
        & 0 
        & 0
        & 0 \\[0.08cm]
    \hline
    & & & \\[-0.25cm]
      0
        & t_{101} t_{110}
        & 0 
        & \sqrt{2} \, t_{110} t_{111}
    \end{array}
  \right)$, \\[0.35cm]
  $\displaystyle
  \left(
    \begin{array}{C{1.70cm}|C{1.70cm}C{1.70cm}|C{1.70cm}}
      0 
        & 0 
        & 0
        & 0 \\[0.08cm]
    \hline
    & & & \\[-0.25cm]
      0 
        & 0 
        & 0
        & 0 \\
      0
        & 0
        & 0 
        & 0 \\[0.08cm]
    \hline
    & & & \\[-0.25cm]
      0 
        & 0 
        & 0
        & 0
    \end{array}
  \right)$, \\[0.35cm]
  $\displaystyle
  \left(
    \begin{array}{C{1.70cm}|C{1.70cm}C{1.70cm}|C{1.70cm}}
      0
        & 0
        & 0 
        & t_{011} t_{011} \\[0.08cm]
    \hline
    & & & \\[-0.25cm]
      0
        & t_{011} t_{101}
        & 0 
        & \sqrt{2} \, t_{011} t_{111} \\
      0 
        & 0 
        & 0
        & 0 \\[0.08cm]
    \hline
    & & & \\[-0.25cm]
      t_{101} t_{101} 
        & \sqrt{2} \, t_{101} t_{111} 
        & 0 
        & t_{111} t_{111}
    \end{array}
  \right)$. \\[0.35cm]
\end{center}

}

\newpage

Finally, let us show the last two $4 \times 4 \times 4$ arrays for the special
case where $f(1,1,1) = t_{111} = 0$. We obtain

{\scriptsize

\begin{center}
  $\displaystyle
  \left(
    \begin{array}{C{1.70cm}|C{1.70cm}C{1.70cm}|C{1.70cm}}
      t_{000} t_{000} 
        & 0
        & 0
        & 0 \\[0.08cm]
    \hline
     & & & \\[-0.25cm]
      0 
        & t_{000} t_{110}
        & 0 
        & 0 \\
      0 
        & 0 
        & t_{000} t_{110}
        & 0 \\[0.08cm]
    \hline
    & & & \\[-0.25cm]
      0
        & 0
        & 0 
        & t_{110} t_{110}
    \end{array}
  \right)$, \\[0.26cm]
  $\displaystyle
  \left(
    \begin{array}{C{1.70cm}|C{1.70cm}C{1.70cm}|C{1.70cm}}
      0
        & t_{000} t_{011}
        & 0 
        & 0 \\[0.08cm]
    \hline
    & & & \\[-0.25cm]
      t_{000} t_{101}
        & 0
        & 0 
        & t_{011} t_{110}  \\
      0 
        & 0 
        & 0
        & 0 \\[0.08cm]
    \hline
    & & & \\[-0.25cm]
      0
        & t_{101} t_{110}
        & 0 
        & 0
    \end{array}
  \right)$, \\[0.26cm]
  $\displaystyle
  \left(
    \begin{array}{C{1.70cm}|C{1.70cm}C{1.70cm}|C{1.70cm}}
      0 
        & 0 
        & t_{000} t_{011} 
        & 0 \\[0.08cm]
    \hline
    & & & \\[-0.25cm]
      0 
        & 0 
        & 0
        & 0 \\
      t_{000} t_{101}
        & 0
        & 0 
        & - t_{011} t_{110} \\[0.08cm]
    \hline
    & & & \\[-0.25cm]
      0 
        & 0 
        & - t_{101} t_{110}
        & 0
    \end{array}
  \right)$, \\[0.26cm]
  $\displaystyle
  \left(
    \begin{array}{C{1.70cm}|C{1.70cm}C{1.70cm}|C{1.70cm}}
      0
        & 0
        & 0 
        & t_{011} t_{011} \\[0.08cm]
    \hline
    & & & \\[-0.25cm]
      0
        & t_{011} t_{101}
        & 0 
        & 0 \\
      0 
        & 0 
        & - t_{011} t_{101}
        & 0 \\[0.08cm]
    \hline
    & & & \\[-0.25cm]
      t_{101} t_{101} 
        & 0
        & 0 
        & 0
    \end{array}
  \right)$ \\[0.26cm]
\end{center}

}

\noindent
and

{\scriptsize

\begin{center}
  $\displaystyle
  \left(
    \begin{array}{C{1.70cm}|C{1.70cm}C{1.70cm}|C{1.70cm}}
      t_{000} t_{000} 
        & 0
        & 0
        & 0 \\[0.08cm]
    \hline
     & & & \\[-0.25cm]
      0 
        & t_{000} t_{110}
        & 0 
        & 0 \\
      0 
        & 0 
        & 0
        & 0 \\[0.08cm]
    \hline
    & & & \\[-0.25cm]
      0
        & 0
        & 0 
        & t_{110} t_{110}
    \end{array}
  \right)$, \\[0.26cm]
  $\displaystyle
  \left(
    \begin{array}{C{1.70cm}|C{1.70cm}C{1.70cm}|C{1.70cm}}
      0
        & t_{000} t_{011}
        & 0 
        & 0 \\[0.08cm]
    \hline
    & & & \\[-0.25cm]
      t_{000} t_{101}
        & 0
        & 0 
        & t_{011} t_{110}  \\
      0 
        & 0 
        & 0
        & 0 \\[0.08cm]
    \hline
    & & & \\[-0.25cm]
      0
        & t_{101} t_{110}
        & 0 
        & 0
    \end{array}
  \right)$, \\[0.26cm]
  $\displaystyle
  \left(
    \begin{array}{C{1.70cm}|C{1.70cm}C{1.70cm}|C{1.70cm}}
      0 
        & 0 
        & 0
        & 0 \\[0.08cm]
    \hline
    & & & \\[-0.25cm]
      0 
        & 0 
        & 0
        & 0 \\
      0
        & 0
        & 0 
        & 0 \\[0.08cm]
    \hline
    & & & \\[-0.25cm]
      0 
        & 0 
        & 0
        & 0
    \end{array}
  \right)$, \\[0.26cm]
  $\displaystyle
  \left(
    \begin{array}{C{1.70cm}|C{1.70cm}C{1.70cm}|C{1.70cm}}
      0
        & 0
        & 0 
        & t_{011} t_{011} \\[0.08cm]
    \hline
    & & & \\[-0.25cm]
      0
        & t_{011} t_{101}
        & 0 
        & 0 \\
      0 
        & 0 
        & 0
        & 0 \\[0.08cm]
    \hline
    & & & \\[-0.25cm]
      t_{101} t_{101} 
        & 0
        & 0 
        & 0
    \end{array}
  \right)$. \\[0.26cm]
\end{center}

}

In particular, for the function $\fzLCT$ in Lemma~\ref{lemma:f:LCT:1} we obtain

{\scriptsize

\begin{center}
  $\displaystyle
  \left(
    \begin{array}{C{1.70cm}|C{1.70cm}C{1.70cm}|C{1.70cm}}
      \alpha^2
        & 0
        & 0
        & 0 \\[0.08cm]
    \hline
     & & & \\[-0.25cm]
      0 
        & \alpha \beta
        & 0 
        & 0 \\
      0 
        & 0 
        & 0
        & 0 \\[0.08cm]
    \hline
    & & & \\[-0.25cm]
      0
        & 0
        & 0 
        & \beta^2
    \end{array}
  \right)$, \\[0.26cm]
  $\displaystyle
  \left(
    \begin{array}{C{1.70cm}|C{1.70cm}C{1.70cm}|C{1.70cm}}
      0
        & \alpha \beta
        & 0 
        & 0 \\[0.08cm]
    \hline
    & & & \\[-0.25cm]
      \alpha \beta
        & 0
        & 0 
        & \beta^2 \\
      0 
        & 0 
        & 0
        & 0 \\[0.08cm]
    \hline
    & & & \\[-0.25cm]
      0
        & \beta^2
        & 0 
        & 0
    \end{array}
  \right)$, \\[0.26cm]
  $\displaystyle
  \left(
    \begin{array}{C{1.70cm}|C{1.70cm}C{1.70cm}|C{1.70cm}}
      0 
        & 0 
        & 0
        & 0 \\[0.08cm]
    \hline
    & & & \\[-0.25cm]
      0 
        & 0 
        & 0
        & 0 \\
      0
        & 0
        & 0 
        & 0 \\[0.08cm]
    \hline
    & & & \\[-0.25cm]
      0 
        & 0 
        & 0
        & 0
    \end{array}
  \right)$, \\[0.26cm]
  $\displaystyle
  \left(
    \begin{array}{C{1.70cm}|C{1.70cm}C{1.70cm}|C{1.70cm}}
      0
        & 0
        & 0 
        & \beta^2 \\[0.08cm]
    \hline
    & & & \\[-0.25cm]
      0
        & \beta^2
        & 0 
        & 0 \\
      0 
        & 0 
        & 0
        & 0 \\[0.08cm]
    \hline
    & & & \\[-0.25cm]
      \beta^2
        & 0
        & 0 
        & 0
    \end{array}
  \right)$, \\[0.26cm]
\end{center}

}

\noindent
where $\alpha \defeq \frac{2 + 6\theta}{\sqrt{8}}$ and
$\beta \defeq \frac{2 - 2\theta}{\sqrt{8}}$. Analyzing the support of the
function $\hat{f}$, yields some of the results mentioned in
Section~\ref{sec:dct:1}.

\section{Incidence Matrix used in Section~\ref{sec:numerical:results:1}}
  \label{app:incidence:matrix:1}

The incidence matrix used in Section~\ref{sec:numerical:results:1} is
\begin{align*}
  \matr{H}
    &= \left(
         \begin{array}{cccccccccccc}
           0 & 0 & 0 & 0 & 0 & 0 & 0 & 0 & 1 & 1 & 0 & 1 \\
           0 & 0 & 1 & 0 & 0 & 0 & 0 & 1 & 0 & 0 & 1 & 0 \\
           0 & 0 & 1 & 0 & 0 & 1 & 0 & 0 & 1 & 0 & 0 & 0 \\
           0 & 0 & 0 & 0 & 1 & 1 & 1 & 0 & 0 & 0 & 0 & 0 \\
           1 & 1 & 0 & 0 & 1 & 0 & 0 & 0 & 0 & 0 & 0 & 0 \\
           1 & 0 & 0 & 0 & 0 & 0 & 0 & 0 & 0 & 0 & 1 & 1 \\
           0 & 0 & 0 & 1 & 0 & 0 & 0 & 1 & 0 & 1 & 0 & 0 \\
           0 & 1 & 0 & 1 & 0 & 0 & 1 & 0 & 0 & 0 & 0 & 0
         \end{array}
       \right).
\end{align*}

\vspace{0.25cm}

\IEEEtriggeratref{7}

\bibliographystyle{IEEEtran}
\bibliography{/home/vontobel/references/references.bib,refs.bib}

\end{document}